\documentclass[a4paper]{article}

\usepackage{icrc2013}
\usepackage[english]{babel}
\usepackage{overpic}
\usepackage{amssymb}

\def\degr{\hbox{$^\circ$}}
\def\arcmin{\hbox{$^\prime$}}
\def\arcsec{\hbox{$^{\prime\prime}$}}

\newcommand{\linebreakcell}[2][c]{%
  \begin{tabular}[#1]{@{}c@{}}#2\end{tabular}}

\title{Search for Very-High-Energy Gamma-Ray Emission from Young Supernovae with H.E.S.S.}

\shorttitle{Search for VHE Gamma-Ray Emission from Young SNe with H.E.S.S.}

\authors{
Dirk Lennarz$^{1,2}$
for the H.E.S.S. collaboration
}

\afiliations{
$^1$ Max-Planck-Institut f\"ur Kernphysik, P.O. Box 103980, D 69029 Heidelberg, Germany \\
\scriptsize{
$^2$ now at: School of Physics and Center for Relativistic Astrophysics, Georgia Institute of Technology, Atlanta, Georgia, USA \\
}
}

\email{dirk.lennarz@gatech.edu}

\abstract{Supernova (SN) remnants are a well motivated candidate for the acceleration sites of cosmic rays with energies up to the knee ($10^{15}$~eV). It has been suggested that also young SNe ($\lesssim 1$~year after the explosion) may be able to accelerate cosmic rays to even higher energies. A smoking gun for cosmic-ray acceleration in young SNe would be the production of very-high-energy (VHE, $>100$~GeV) gamma-ray radiation. The H.E.S.S. imaging air Cherenkov telescope array is an instrument sensitive to such radiation. In this contribution, the pointing directions of the H.E.S.S. telescopes are compared to a recently published, extragalactic SN catalogue to identify coincidental observations. The results of the data analysis are discussed.}

\keywords{H.E.S.S., supernovae, very-high energy, gamma rays}

\begin{document}
\maketitle

\section{Introduction}
A core collapse at the end of the lifetime of a massive star leads to a supernova explosion (SN). The idea that those events might be connected to cosmic rays was suggested very early \cite{bib:SNe_CRs} and can be convincingly justified from the SN energetics \cite{bib:origin_of_cosmic_rays}: an acceleration efficiency of $\sim1$\% would already be enough to sustain the cosmic ray energy density in our Galaxy. Diffusive shock acceleration is an efficient acceleration mechanism for cosmic rays in SN remnants (SNRs). SNRs may be able to accelerate protons up to energies in the range of $10^{15}$~eV, but it remains unclear if that is sufficient to explain all Galactic cosmic rays up to the \emph{ankle} around $10^{18.5}$~eV.

There is reason to believe that ``young'' SNe ($\lesssim 1$~year after the explosion) might be able to accelerate cosmic rays to energies higher than $10^{15}$~eV. A possible scenario includes the energy release of a compact object (e.g. a pulsar) that forms as a result of a core-collapse SN which can accelerate particles with great efficiency to relativistic energies (internal acceleration). After the SN shock has left the stellar envelope, it starts to interact with the circum-stellar material and continues as an external shock. In the external acceleration scenario, the compression of the gas at the shock front creates a reverse (or internal) shock and diffusive shock acceleration might be more efficient in a thin shell between the two shock fronts than at a single shock. This scenario has very successfully reproduced the radio emission of several SNe \cite{bib:shockmodel_3,bib:shockmodel_4}.

It has been realised that in both scenarios very-high-energy (VHE, $>100$ GeV) gamma-ray and neutrino radiation might be used as evidence for particle acceleration \cite{bib:pulsarmodel_1,bib:shockmodel_1}. The characteristic time-scale of the emission is one year after the explosion. There has been a search for TeV neutrino radiation with data from the AMANDA neutrino telescope \cite{bib:ICRC_lennarz,bib:diploma_lennarz}. No indication of a signal has been seen and the limits lie at the upper end of the theoretically motivated parameter range and thus only marginally constrain the assumed model.

\section{H.E.S.S. Supernova Observations}\label{sec:catalogue}
\begin{table*}[t!]
\begin{center}
\setlength{\tabcolsep}{4pt}
\begin{tabular}{ccccccc}
  \hline
  SN & Host galaxy & RA J2000 & DEC J2000 & Dist. [Mpc] & Type & Disc. date\\
  \hline
  2004cc & NGC 4568 & $12^{\rm h}36^{\rm m}34.40^{\rm s}$ & $+11\degr14\arcmin32.8\arcsec$ & $25\pm3$ & Ic & 2004-06-10\\
  2004cx & NGC 7755 & $23^{\rm h}47^{\rm m}52.86^{\rm s}$ & $-30\degr31\arcmin32.6\arcsec$ & $26\pm5$ & II & 2004-06-26\\
  2004gk & IC 3311  & $12^{\rm h}25^{\rm m}33.21^{\rm a}$ & $+12\degr15\arcmin39.9\arcsec$ & $17\pm3$ & Ic & 2004-11-25\\
  2004gn & NGC 4527 & $12^{\rm h}34^{\rm m}12.10^{\rm s}$ & $+02\degr39\arcmin34.4\arcsec$ & $12.6\pm0.5$ & Ic & 2004-12-01\\
  2006mr & NGC 1316 & $03^{\rm h}22^{\rm m}42.84^{\rm s}$ & $-37\degr12\arcmin28.5\arcsec$ & $12.6\pm0.6$ & Ia & 2006-11-05\\
  2007cj & IC 2531  & $09^{\rm h}59^{\rm m}55.76^{\rm s}$ & $-29\degr37\arcmin03.3\arcsec$ & $29\pm6$ & Ia & 2007-05-03\\
  2008bk & NGC 7793 & $23^{\rm h}57^{\rm m}50.42^{\rm s}$ & $-32\degr33\arcmin21.5\arcsec$ & $4.0\pm0.4$ & IIP? & 2008-03-25\\
  2008bp & NGC 3095 & $10^{\rm h}00^{\rm m}01.57^{\rm s}$ & $-31\degr33\arcmin21.8\arcsec$ & $29\pm6$ & IIP & 2008-04-02\\
  2009js & NGC 918  & $02^{\rm h}25^{\rm m}48.28^{\rm s}$ & $+18\degr29\arcmin25.8\arcsec$ & $16\pm3$ & IIP? & 2009-10-11\\
  \hline
\end{tabular}
\caption{List of nearby SNe serendipitously observed by the H.E.S.S. telescopes and relevant data from the unified SN catalogue \cite{bib:my_SN_catalogue}. A type followed by a question mark means that the SN type is uncertain.}
\label{tab:serendipitous_SNe}
\end{center}
\end{table*}
\subsection{The High Energy Stereoscopic System}
H.E.S.S., located 1800~m above sea level in the Khomas Highland of Namibia, consists of four Imaging Atmospheric Cherenkov Telescopes  \cite{bib:HESS_crab}. It is sensitive to VHE gamma rays between hundreds of GeV to tens of TeV by detecting Cherenkov light emitted when a gamma ray is absorbed in the atmosphere in an extensive air shower. Each telescope has $\sim100~\rm{m^2}$ tessellated mirror surface and features a pixelated camera of 960 photomultiplier tubes (PMTs). Each pixel corresponds to approximately $0.16\degr$ of the sky, resulting in a total field of view of $5\degr$ in diameter. Due to the large field of view, young SNe are observed serendipitously during the observation of other targets. The angular resolution (68\% containment) of H.E.S.S. is typically $0.1\degr$ and the energy resolution $\sim15\%$.

\subsection{A Unified Supernova Catalogue}
Suitable H.E.S.S. observations, with a SN in the field of view, within one year after the SN discovery, are selected using a newly published SN catalogue \cite{bib:my_SN_catalogue}. This new and unified catalogue includes journal-refereed distances to the host galaxies and therefore allows to replace the redshift based estimate of the distance, which is especially bad for nearby SNe that are most interesting for the current analysis. Furthermore, the unified SN catalogue tries to resolve inconsistencies in the listed information, which enables its use as a meta-catalogue of the current SN catalogues. A subset of high-quality SNe with more reliable information can easily be selected with the meta-data.

H.E.S.S. observations are selected based on the following selection criteria:
\begin{itemize}
   \item SN within 2.5\degr~of the nominal H.E.S.S. pointing direction,
   \item observation takes place within one year after the SN discovery date,
   \item SN with known redshift $z$ and $z < 0.01$.
\end{itemize}
The angular distance cut is introduced because of possible systematic problems of the H.E.S.S. data analysis at larger offsets. It has recently been shown that an average offset of 2.0\degr~does not introduce a significant bias \cite{bib:1ES_1312-423}. The time cut can e.g. be motivated from some simple, analytical modelling, showing that most of the emission can be expected during the first year \cite{bib:pulsarmodel_3}. Since the expected flux for an extragalactic SN is low, the distance cut restricts the search to nearby SNe only.

Table~\ref{tab:serendipitous_SNe} contains all SNe found with the search criteria and the relevant information as listed in the unified SN catalogue. Strikingly, the closest SN has a distance below 5~Mpc. The list also contains two SNe of type Ia. According to the acceleration mechanism discussed before, no VHE emission can be expected from them, because SNe of type Ia do not produce a compact object or are expected to have a strong reverse shock due to the lack of circumstellar material. However, the two SNe are investigated as well to explore the possibility of unexpected results.

Radio SNe might indicate that the external acceleration is particular powerful and are therefore prime targets to search for VHE emission. However, only a handful of radio SNe have been observed to date and for most SNe the information whether or not it is a radio SN is not available.

\subsection{Analysis}
The data calibration, image cleaning, Hillas moment calculation and event reconstruction \cite{bib:HESS_crab} is done using the standard H.E.S.S. analysis software\footnote{version hap-12-06-pl03}. The background caused by cosmic ray showers is rejected using the cuts for the ``hard'' configuration \cite{bib:HESS_crab}, enhanced by boosted decision trees \cite{bib:TMVA}. This tight selection is used because the expected fluxes for extragalactic SNe are low. It also determine the optimal size of the signal region (``on-region'') around the SN position. A background estimate is obtained from regions in the same field of view (``off-regions'') using the reflected-region-background model \cite{bib:BG_paper}.

\subsection{Results}
\begin{table*}[t!]
\begin{center}
\setlength{\tabcolsep}{4pt}
\begin{tabular}{ccccccccc}
  \hline
  SN & Exp. $[h]$ & Offset $[^{\rm o}]$ & Delay $[d]$ & $N_{\rm{on}}$ & $N_{\rm{off}}$ & $\alpha$ & $N_{\rm excess}$ & Significance \\
  \hline
  2004cc & 26.0 & 1.9 & 287 & 79  & 3860 & 0.019 &   4  &  0.5\\
  2004cx & 68.8 & 2.2 & 176 & 85  & 4237 & 0.015 &  20  &  2.4\\
  2004gk & 26.0 & 1.4 & 119 & 114 & 3764 & 0.027 &  11  &  1.1\\
  2004gn &  7.5 & 1.5 & 127 & 44  & 1175 & 0.024 &  15  &  2.6\\
  2006mr & 17.8 & 1.2 & 320 & 24  & 669  & 0.043 &  -5  & -0.9\\
  2007cj &  4.3 & 2.4 & 143 & 3   & 133  & 0.014 &   1  &  ---\\
  2008bk & 10.7 & 1.9 & 142 & 14  & 588  & 0.020 &   3  &  0.7\\
  2008bp &  4.8 & 2.1 & 277 & 4   & 222  & 0.017 &  0.3 &  ---\\
  2009js &  3.8 & 2.1 &  14 & 8   & 359  & 0.016 &   2  &  ---\\
  \hline
\end{tabular}
\caption{Preliminary results of the search for excess photons for the H.E.S.S. SN observations. The column ``Exp.'' gives the deadtime corrected livetime of the used observations, the columns ``Offset'' and ``Delay'' the average offset of the SN from the nominal observation position and the average number of days between the observations and the discovery date of the SN. $N_{\rm{on}}$ is the number of gamma-ray candidates in the signal region around the SN position and $N_{\rm{off}}$ the background estimate. When scaled by the normalisation factor $\alpha$ they yield the number of excess events $N_{\rm excess} = N_{\rm{on}}-\alpha N_{\rm{off}}$. The significance is estimated using Eq. (17) of \cite{bib:LiMa}, which is only applicable when there are at least 10 counts in the signal region.}
\label{tab:SNe_results_detection}
\end{center}
\end{table*}

\begin{figure}[t!]
 \centering
 \begin{overpic}[width=0.49\textwidth]{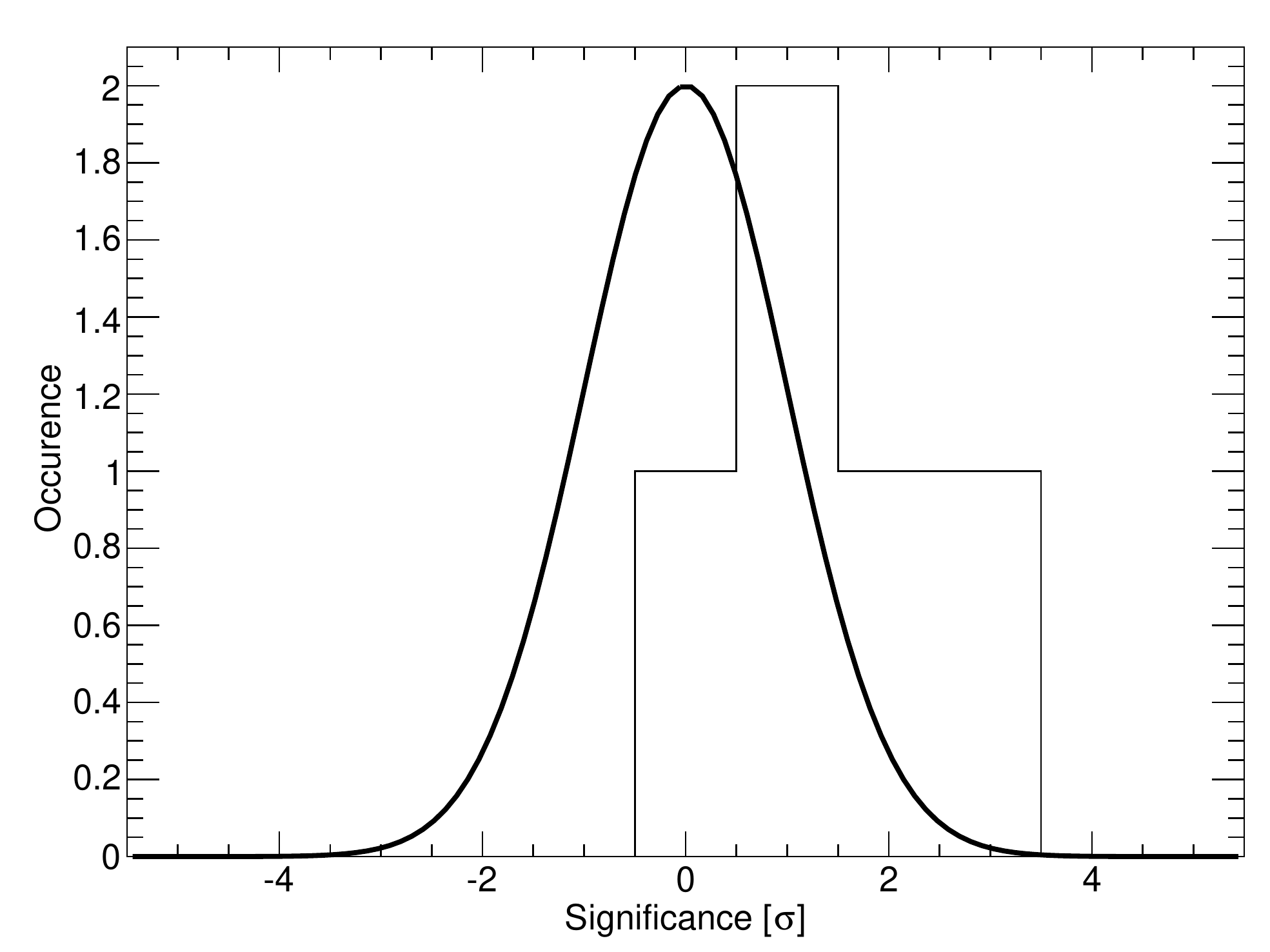}
 \put(16,62){\LARGE Preliminary}
 \end{overpic}
 \caption{Significance distribution obtained from the number of excess photons for the H.E.S.S. SN observations of core-collapse SNe (excluding SN~2006mr). To guide the eye a Gaussian with mean zero and width one is added.}
 \label{fig:SN_significance_distribution}
\end{figure}

Table~\ref{tab:SNe_results_detection} shows the preliminary results of the analysis of all available H.E.S.S. data with good hardware status. No significant emission from any individual SN is observed. It is striking that except for SN~2006mr, which is a type Ia SN and therefore no VHE emission is expected, all SNe have a positive significance. Figure~\ref{fig:SN_significance_distribution} shows the significance distribution for the core-collapse SNe and a comparison to a Gaussian with mean zero and width one indicates a slight offset between the two. The individual core-collapse SNe have also been stacked, using also those where the significance cannot be calculated. This analysis yields a significance of 3.2, which hints at the indication of emission. A large contribution to the possible signal comes from SN~2004cx and SN~2004gn.

The results have been crosschecked with an independent calibration and analysis of the data \cite{bib:model}. In the cross-check analysis the significance for SN~2004cx and SN~2004gn is 0.2 and -0.6, respectively. Furthermore, the results have also been cross-checked using the H.E.S.S. standards cuts enhanced by boosted decision trees, which should be less sensitive than the hard cuts. In the cross-check analysis the significance drops to 2.2 for SN~2004cx and -0.3 for SN~2004gn. Stacking the results of the analysis of all SNe with standards cuts yields a total significance of 1.5.

Given the results of the cross-check analyses it appears that the indication of emission seen in the current analysis is most likely a statistical fluctuation and the results are still compatible with the assumption of no signal. Following the individual non-detections, upper limits on the fluxes are calculated for an $E^{-2}$ spectrum (see Table~\ref{tab:SNe_results_spectral}). For the H.E.S.S. spectral analysis only observations taking during good weather conditions are used to avoid possible biases in the energy reconstruction.

\begin{table}[t!]
\begin{center}
\setlength{\tabcolsep}{3pt}
\begin{tabular}{cccccc}
  \hline
  SN & \linebreakcell[t]{Exp. \\ $[$h]} & \linebreakcell[t]{Offset \\ $[\deg]$} & \linebreakcell[t]{Delay \\ $[$d]} & \linebreakcell[t]{$E_{\rm th}$ \\ $[$TeV]} & \linebreakcell[t]{Flux ULs $>1$~TeV \\ $[10^{-13}\rm cm^{-2}s^{-1}]$}\\
  \hline
  2004cc & 21.8 & 1.9 & 288 & 0.62 &  1.8 \\
  2004cx & 42.1 & 2.3 & 197 & 0.42 &  2.3 \\
  2004gk & 21.8 & 1.4 & 120 & 0.62 &  1.7 \\
  2004gn &  7.5 & 1.5 & 127 & 0.46 &  4.1 \\
  2006mr &  2.0 & 2.0 & 334 & 0.51 &  4.9 \\
  2007cj &  4.3 & 2.4 & 143 & 0.62 & 12.0 \\
  2008bk &  8.1 & 1.9 & 136 & 0.51 &  4.2 \\
  2008bp &  4.1 & 2.1 & 276 & 0.46 &  7.9 \\
  2009js &  3.8 & 2.1 &  14 & 1.00 & 12.0 \\
  \hline
\end{tabular}
\caption{Preliminary integral flux upper limits for a confidence level of 95\% derived from the H.E.S.S. spectral analysis for an $E^{-2}$ spectrum together with the energy threshold ($E_{\rm th}$). The first three columns are the same as in Table~\ref{tab:SNe_results_detection}, but now for the observations taken under good weather conditions.}
\label{tab:SNe_results_spectral}
\end{center}
\end{table}

\section{Summary}\label{sec:summary}
The current analysis shows no significant signs of emission for extragalactic SNe observed serendipitously by the H.E.S.S. telescopes, so young SNe cannot be unexpectedly strong emitters of VHE gamma-ray radiation. The derived limits can be used to derive astrophysical implications for modelling, which are not explored here. However, definite conclusions would need further observational input like e.g. a detection and measurements of the pulsar properties inside the SN remnant or the properties of the circumstellar material. The next generation instruments like CTA, with their improved sensitivity, will have better chances to detect VHE emission from extragalactic, young SNe. For a Galactic SN even the current generation of Cherenkov telescopes might be able to confirme or rule out some of the proposed modelling.

\vspace*{0.5cm}
\footnotesize{{\bf Acknowledgment:}{The support of the Namibian authorities and of the University of Namibia in facilitating the construction and operation of H.E.S.S. is gratefully acknowledged, as is the support by the German Ministry for Education and Research (BMBF), the Max Planck Society, the French Ministry for Research, the CNRS-IN2P3 and the Astroparticle Interdisciplinary Programme of the CNRS, the U.K. Science and Technology Facilities Council (STFC), the IPNP of the Charles University, the Czech Science Foundation, the Polish Ministry of Science and  Higher Education, the South African Department of Science and Technology and National Research Foundation, and by the University of Namibia. We appreciate the excellent work of the technical support staff in Berlin, Durham, Hamburg, Heidelberg, Palaiseau, Paris, Saclay, and in Namibia in the construction and operation of the equipment.}}

\end{document}